\documentclass[prd,twocolumn,nofootinbib,aps,superscriptaddress,tightenlines]{revtex4}
\usepackage{latexsym}
\usepackage{graphicx}
\usepackage{caption2}
\usepackage{psfrag}
\usepackage{amsmath}
\usepackage{supertabular}
\usepackage{longtable}

\def\nn{\nonumber \\ }

\begin{document}

\preprint{ \vbox{\hbox{IFIC/08-05} \hbox{FTUV-07-0129}  } 
}

\title{Novel patterns for vector mesons from the large-$N_c$ limit} 

\preprint{ \vbox{\hbox{IFIC/08-05} \hbox{FTUV-07-0129}  } 
}

\author{Oscar Cat\`a}

\affiliation{Ernest Orlando Lawrence Berkeley National Laboratory, University of California, Berkeley, CA 94720}

\author{Vicent Mateu}

\affiliation{Departament de F\'\i sica Te\`orica, IFIC, Universitat de Val\`encia-CSIC, Apt. Correus 22085, E-46071 Val\`encia, Spain}

\vskip-2.1in
\hskip 6.0in
{\vbox{ \hbox{IFIC/08-05} \hbox{FTUV-07-0129}}}
\vskip0.1in

\begin{abstract}
\noindent We report on a relation between the decay constants of $\rho$-like $J^{PC}=1^{--}$ vector mesons, which arises solely from the perturbative analysis of the $\langle VV\rangle$, $\langle TT\rangle$ and $\langle VT\rangle$ correlators at $\mathcal{O}(\alpha_s^0)$ in the large-$N_c$ limit. We find $f_{V}^T/f_{V}=1/\sqrt{2}$ for highly excited states together with a pattern of alternation in sign. Quite remarkably, recent lattice determinations reported $f_{\rho}^T/f_{\rho}=0.72(2)$, in excellent agreement with our large-$N_c$ result. This seems to suggest a pattern like $f_{Vn}^T/f_{Vn}=(-1)^n/\sqrt{2}$ for the whole $(1^{--})$ states. In order to test this conjecture in real QCD we construct a set of spectral sum rules, which turn out to comply nicely with this scenario. 
\end{abstract}

\date{January 2008}
\pacs{77}
\maketitle

\section{Introduction}
\label{sec:intro}

\noindent QCD in the $1/N_c$ expansion~\cite{'tHooft:73} is still nowadays the most promising analytical tool to deal with the strong interactions in the non-perturbative regime. Its success in providing, already at leading order, a satisfactory explanation for the OZI rule, the suppression of exotics in the meson spectrum or the dominance of one-particle over multi-particle states in virtual resonance exchange processes is seen as strong indication not only that the $1/N_c$ expansion is capturing the qualitative features of QCD, but also that its degree of convergence is remarkable, at least for certain observables.

Unfortunately, in spite of numerous efforts, no Lagrangian formulation of large-$N_c$ QCD below the confining scale is known.\footnote{For certain phenomenological applications, the formulation of~\cite{Ecker} has proven very successful, but one has to truncate the spectrum and assume lowest meson dominance.} However, a great deal of information about the theory at leading order can be extracted from a general analysis of its correlators~\cite{Witten:79}. Assuming that confinement persists for arbitrarily large number of colors, planar Feynman diagrams at the quark-gluon level give rise to a dual hadron picture. 

The emerging picture of large-$N_c$ QCD as a theory of free, stable, non-interacting mesons seems quite far away from what we observe in the QCD spectrum. However, the picture turns out to be extremely accurate for space-like momenta. Thanks to this observation~\cite{rev}, in the last decade large-$N_c$ QCD experienced a considerable boost as a tool in the quantitative understanding of electroweak observables, such as $(g-2)_{\mu}$ or $B_K$~\cite{works}.

Despite its successes in explaining many patterns of QCD, the understanding of the theory, even in the strict large-$N_c$ limit, is still rather limited. For instance, even though the analytic structure of the correlators is known to consist of an array of single pole singularities, the location of the poles ({\it{i.e.}} the resonance masses) and their residues (decay constants) cannot be computed. One usually assumes that their values cannot differ much from the experimentally measured ones, allowing for a na\"\i ve systematic 30\% uncertainty. 

In this work we want to report on a quantitative prediction of the large-$N_c$ limit in the sector of light-flavored vector mesons. In particular, we will show that perturbative QCD alone sets a relation between the couplings of vector mesons to the vector (${\bar{q}}\,\gamma_{\mu} q$) and 
tensor current (${\bar{q}}\,\sigma_{\mu\nu} q$). 

This power of prediction is due to the exceptional status of the two-point correlators $\Pi_{VV}$, $\Pi_{TT}$ and $\Pi_{VT}$ (to be defined in the next section). $J^{PC}=1^{--}$ $\rho$-like mesons are exchanged in the three correlators, a situation that strongly constrains and, as we will show, sets a distinct pattern for the decay constants of vector mesons in the large-$N_c$ limit. To the best of our knowledge, no similar self-constrained set of correlators exists for particles other than vector mesons. This system of correlators and the need to consider them simultaneously was first discussed in Ref.~\cite{Craigie:81} in the context of QCD sum rules.

Obviously, it is an interesting issue to find out how stable our prediction is when one moves to real QCD. The quantity $f_{\rho}^T/f_{\rho}$ was computed recently in lattice QCD~\cite{Bec,Braun:2003jg,Donnellan:2007xr}. Interestingly, the value reported is in excellent agreement with our prediction in the large-$N_c$ limit. Therefore, the possibility that $f_{Vn}^T/f_{Vn}$ be approximately constant in the QCD spectrum is very suggestive and is investigated in the second part of the paper, where we will show that sum rules are indeed compatible with this scenario.

This paper is organized as follows\,: in Section~\ref{sec:def} we set our notation and define the relevant correlators. Their expressions at large-$N_c$ are worked out in Section~\ref{sec:large}. In Section~\ref{sec:pred} we derive the prediction on the ratio of decay rates $f_{Vn}^T/f_{Vn}$. This prediction is then tested at low energies by using a set of finite energy sum rules. This is done in Section~\ref{sec:sumrules}. Finally, Section~\ref{sec:disc} contains our conclusions and outlook for future work.
\section{Definitions}
\label{sec:def}

\noindent Let us begin by defining the following set of two-point correlators
\begin{eqnarray}
\Pi^{VV}_{\mu\nu}(q)&=&i\int\mathrm{d}^{4}x\, e^{iq\cdot x}\langle \,0\,|\,T\lbrace\, V_{\mu}(x)\,V_{\nu}^{\dagger}(0)\,\rbrace |\,0\,\rangle\, , \nonumber\\
\Pi^{VT}_{\mu;\nu\rho}(q)&=&i\int\mathrm{d}^{4}x\, e^{iq\cdot x}\langle\, 0\,|\,T\lbrace\, V_{\mu}(x)\,T_{\nu\rho}^{\dagger}(0)\,\rbrace |\,0\,\rangle \, , \nonumber\\
\Pi^{TT}_{\mu\nu;\alpha\beta}(q)&=&i\int\mathrm{d}^{4}x\, e^{iq\cdot x}\langle \,0\,|\,T\lbrace\, T_{\mu\nu}(x)\,T_{\alpha\beta}^{\dagger}(0)\,\rbrace |\,0\,\rangle\, , \nonumber\\
\end{eqnarray}
where $V_{\mu}(x)$ and $T_{\mu\nu}(x)$ stand for the QCD color singlet currents 
\begin{eqnarray}
T_{\mu\nu}(x)&=&:\bar{u}(x)\sigma_{\mu\nu}d(x):\, ,\nonumber\\
V_{\mu}(x)&=&:\bar{u}(x)\gamma_{\mu}d(x):\,,
\end{eqnarray}
and $\sigma^{\mu\nu}=i/2\,[\gamma^{\mu},\gamma^{\nu}]$. In the chiral limit, 
vector current conservation, Lorentz symmetry, parity conservation and the antisymmetry of the tensor indices imply the following kinematical structures\,:
\begin{eqnarray}\label{ward}
\Pi_{VV}^{\mu\nu}(q)&=&(q^{\mu}q^{\nu}-q^2g^{\mu\nu})\,\Pi_{VV}(q^2) \, , \nonumber\\
\Pi_{VT}^{\mu;\nu\rho}(q)&=&i\,(q^{\rho}g^{\mu\nu}-q^{\nu}g^{\mu\rho})\,\Pi_{VT}(q^2)\, ,
\end{eqnarray}
such that each Green function can be expressed in terms of a single form factor. In contrast, there are two independent kinematical structures for the tensor correlator\,: 
\begin{equation}\label{deftensor}     
\Pi_{TT}^{\mu\nu;\alpha\beta}(q)\,=\, \Pi_{TT}^-(q^2)F_-^{\mu\nu;\alpha\beta}\,+\,\Pi_{TT}^+(q^2)F_+^{\mu\nu;\alpha\beta}\, .
\end{equation}
For phenomenological purposes it is convenient to project the form factors in combinations with well-defined parity, $\Pi_{TT}^{\pm}$. $F_-^{\mu\nu;\alpha\beta}$ and $F_+^{\mu\nu;\alpha\beta}$ are
Lorentz tensors given by 
\begin{eqnarray}\label{structure}
F_-^{\mu\nu;\alpha\beta} & = & q^{\mu}q^{\beta}g^{\nu\alpha}+q^{\nu}q^{\alpha}g^{\mu\beta}-q^{\mu}q^{\alpha}g^{\nu\beta}-q^{\nu}q^{\beta}g^{\mu\alpha}\,,\nonumber\\
F_+^{\mu\nu;\alpha\beta} & = & -\,\varepsilon^{\mu\nu\sigma\rho}\,\varepsilon^{\alpha\beta\gamma\tau}\,g_{\sigma\gamma}\,q_{\rho}\,q_{\tau}\nonumber\\
&=&F_-^{\mu\nu;\alpha\beta}\,+\,q^{2}\left(g^{\mu\alpha}g^{\nu\beta}-g^{\mu\beta}g^{\nu\alpha}\right)\,,
\end{eqnarray}
which project onto the different parity-even and parity-odd sectors of the correlator. It is easy to verify that they are idempotent and orthogonal, 
\begin{eqnarray}
F_{\pm\mu\nu;\alpha\beta}\,F_{\mp}^{\alpha\beta;\sigma\rho} & = & 0\,,\nonumber \\
F_{\pm\mu\nu;\alpha\beta}\,F_{\pm\phantom{\alpha}\sigma\rho}^{\alpha\beta} & = & \pm\, 2\,q^{2}\,F_{\pm\mu\nu;\sigma\rho}\,,
\end{eqnarray}
with the normalization
\begin{equation}
F_{\pm\mu\nu;\alpha\beta}\,F_{\pm}^{\alpha\beta;\mu\nu}\,=\, 12\, q^{4}\,.
\end{equation}
\section{Correlators at large-$N_c$}
\label{sec:large}

\noindent We start by listing the analytic properties of the correlators. Two-point correlators are known to be analytic functions in the entire complex $q^2$ plane except on the physical axis. Use of the Cauchy theorem then leads to the so-called Kallen-Lehman representation
\begin{equation}
\Pi(q^2)=\int_{0}^{\infty}\frac{dt}{t-q^{2}}\frac{1}{\pi}\,\mathrm{Im}\,\Pi(t)+{\cal{P}}(q^2)\, ,
\end{equation}
where ${\cal{P}}(q^2)$ is a polynomial whose degree (number of subtractions) is determined by the behavior of the correlators at large space-like momenta. 

In the case of $\Pi_{TT}$ special care must be exercised. The low energy expansion of this correlator can be shown to develop a singularity as $q^2\rightarrow 0$ whose residue is $\Lambda_3$, according to the conventions of Ref.~\cite{CataMateu07}. The pervading problem with tensor sources is that they are not accessible experimentally. Therefore, $\Lambda_3$ can only be reliably estimated using lattice QCD. We know that the $\Lambda_3$-term cannot be interpreted as a particle, because the singularity is present in both $\Pi_{TT}^+$ and $\Pi_{TT}^-$, which have opposite parity. However, we do not know of any {\it{a priori}} reason why $\Lambda_3$ should vanish. Thus, apart form the physical cut, the correlator has in principle an isolated singularity at the origin that modifies slightly its dispersion relation. This subtlety was ignored in all previous sum rule analyses, {\it{e.g.}}, Ref.~\cite{Craigie:81}. The remaining two-point correlators are free from such singularities.

For large and negative $q^2$, $\Pi_{VV}$ and $\Pi_{TT}$ are given, to first order in $\alpha_s$, by~\cite{Gov}
\begin{eqnarray}\label{opevvtt}
\Pi_{VV}^{\mathrm{OPE}}(q^{2})&=&-\,\frac{N_{c}}{12\,\pi^2}\log\left(\frac{-q^{2}}{\mu^{2}}\right)\nonumber\\
&\,\,&+\,\frac{1}{12\,\pi}\frac{\langle \alpha_s G^{\mu\nu}G_{\mu\nu}\rangle}{q^4}+{\cal{O}}\!\left(\frac{1}{q^6}\right)\!,\nonumber\\
(\Pi_{TT}^{\pm})^{\mathrm{OPE}}(q^{2})&=&-\,\frac{N_{c}}{24\,\pi^2}\log\left(\frac{-q^{2}}{\mu^{2}}\right)\nonumber\\
&\,\,&-\,\frac{1}{24\,\pi}\frac{\langle \alpha_s G^{\mu\nu}G_{\mu\nu}\rangle}{q^4}+{\cal{O}}\left(\frac{1}{q^6}\right)\!,
\end{eqnarray}
where the first line on each equation is the leading perturbative contribution, whereas the second shows the leading condensate of the Operator Product Expansion (OPE). For $\Pi_{VT}$ the perturbative contribution cancels to all orders and we are left with a pure OPE, the first terms of which are~\cite{Bal}
\begin{equation}\label{shortvt} 
\Pi_{VT}^{\mathrm{OPE}}(q^{2})\,=\,\frac{2\,\langle{\bar{\psi}}\psi\rangle}{q^2}+\frac{2\,g_s}{3}\frac{\langle{\bar{\psi}}\,\sigma_{\mu\nu}\,G^{\mu\nu}\psi\rangle}{q^4}\,+\,{\cal{O}}\left(\frac{1}{q^6}\right).
\end{equation}
In view of Eqs.~(\ref{opevvtt}, \ref{shortvt}), it is straightforward to conclude that $\Pi_{VV}$ and $\Pi_{TT}$ obey a once-subtracted dispersion relation,
\begin{eqnarray}\label{disprel}
\Pi_{VV}(q^{2})&=&\int_{0}^{\infty}\frac{dt}{t-q^{2}}\frac{1}{\pi}\,\mathrm{Im}\,\Pi_{VV}(t)\,+\,a_V\,,\nonumber\\
\Pi_{TT}^{\pm}(q^{2})&=& \int_{0}^{\infty}\frac{dt}{t-q^{2}}\frac{1}{\pi}\,\mathrm{Im}\,\Pi_{TT}^{\pm}(t)\,\pm\, \frac{\Lambda_3}{q^2}\,+\,a_T^{\pm}\,,\nonumber\\
\end{eqnarray}
where in the last equation we have included the singularity generated by the $\Lambda_3$-term, while $\Pi_{VT}$ satisfies an unsubtracted one, namely
\begin{equation}
\Pi_{VT}(q^{2})=\int_{0}^{\infty}\frac{dt}{t-q^{2}}\frac{1}{\pi}\,\mathrm{Im}\,\Pi_{VT}(t)\,.
\end{equation}
In the strict large-$N_c$ limit, two-point functions are saturated by the single-particle exchange of an infinite number of stable mesons. Therefore, the spectral functions above take the simple forms
\begin{eqnarray}\label{spectralfunctions}
\frac{1}{\pi}\,\mathrm{Im}\,\Pi_{VV}(t) & = & \sum_n^{\infty}f_{Vn}^{2}\,\delta(t-m_{Vn}^{2})\,,\nonumber \\
\frac{1}{\pi}\,\mathrm{Im}\,\Pi_{TT}^{+}(t) & = & \sum_n^{\infty} f_{Bn}^{2}\,\delta(t-m_{Bn}^{2})\,,\nonumber\\
\frac{1}{\pi}\,\mathrm{Im}\,\Pi_{TT}^{-}(t) & = & \sum_n^{\infty}(f_{Vn}^T)^{2}\,\delta(t-m_{Vn}^{2})\,,\nonumber \\
\frac{1}{\pi}\,\mathrm{Im}\,\Pi_{VT}(t) & = & \sum_n^{\infty} f_{Vn}f_{Vn}^T m_{Vn}\,\delta(t-m_{Vn}^{2})\, ,
\end{eqnarray}
where the particle content is fixed by quantum numbers. Thus, $m_{Vn}$ refers to the $J^{PC}=1^{--}$ vector mesons, whose first representative is the $\rho(770)$, while $m_{Bn}$ corresponds to the $J^{PC}=1^{+-}$ mesons, such as $b_1(1230)$ and its radial excitations. Note that indeed the resonances exchanged by $\Pi_{TT}^+$ and $\Pi_{TT}^-$ are parity-even and parity-odd, respectively.

The following conventions have been adopted for the one-particle to vacuum matrix elements ($\varepsilon^{0123}=+1$)\,: 
\begin{eqnarray}\label{elements}
\langle\,0\,|\, V_{\mu}(0)\,|\rho_{n}(p,\lambda)\rangle & \dot{=} & f_{Vn}\,m_{Vn}\,\epsilon_{\mu}^{(\lambda)}\,,\nonumber\\
\langle\,0\,|\, T_{\nu\rho}(0)\,|\rho_{n}(p,\lambda)\rangle & \dot{=} & i\,f_{Vn}^{T}(\epsilon_{\nu}^{(\lambda)}\,p_{\rho}-\epsilon_{\rho}^{(\lambda)}\,p_{\nu})\,,\nonumber\\
\langle\,0\,|\, T_{\nu\rho}(0)\,|B_{n}(p,\lambda)\rangle & \dot{=} & i\,f_{Bn}^{T}\,\varepsilon_{\nu\rho\mu\sigma}\,\epsilon_{(\lambda)}^{\mu}\,p^{\sigma}\,.\end{eqnarray}
For future convenience we introduce the parameter $\xi_{n}$, defined as
\begin{equation}
\xi_{n}\,=\,\frac{f_{Vn}^{T}}{f_{Vn}}\,.
\end{equation}
If we plug Eqs.~(\ref{spectralfunctions}) into the dispersion relations above,
we find
\begin{eqnarray}\label{dispexp}
\Pi_{VV}(t) & = & \sum_{n}^{\infty}\frac{f_{Vn}^{2}}{-\,q^{2}+m_{Vn}^{2}}\,,\nonumber \\
\Pi_{TT}^{+}(q^{2}) & = & \sum_{n}^{\infty}\frac{f_{Bn}^{2}}{-\,q^{2}+m_{Bn}^{2}}\,+\,\frac{\Lambda_3}{q^2}\,,\nonumber\\
\Pi_{TT}^{-}(q^{2}) & = & \sum_{n}^{\infty}\xi_{n}^{2}\frac{f_{Vn}^{2}}{-\,q^{2}+m_{Vn}^{2}}\,-\,\frac{\Lambda_3}{q^2}\,,\nonumber \\
\Pi_{VT}(t) & = & \sum_{n}^{\infty}\xi_{n}\frac{f_{Vn}^{2}\,m_{Vn}}{-\,q^{2}+m_{Vn}^{2}}\, ,
\end{eqnarray}
up to subtractions. The previous equations would provide the solution to QCD in the large-$N_c$ limit, should decay constants and masses be determined. However, thus far no attempt to solve the theory has been successful. Therefore, Eqs.~(\ref{dispexp}) contain an infinite number of undetermined parameters, and apparently one is forced to resort to approximations to have some predictive power. The purpose of this paper is to show that Eqs.~(\ref{dispexp}) are so strongly correlated that even without solving large-$N_c$ QCD exact predictions can be extracted. 
    
\section{A large-$N_c$ prediction for $f_V^T/f_V$}
\label{sec:pred}

\noindent In order to derive our result, we will only assume that QCD at large-$N_c$ undergoes a confining phase, such that the quark-gluon picture at large-$N_c$ is dual to a theory of mesons. Note that this is the same assumption adopted in deriving the general features of the theory~\cite{Witten:79}. Therefore, in some sense our paper provides an example of a set of correlators in which large-$N_c$ predictions can be made quantitative for hadronic parameters. 

One of the key results of QCD in the large-$N_c$ limit is that the theory contains an infinite number of stable states. Therefore, we can trade the infinite sums above for integrals over the resonance counting index $n$ with the use of Euler-Maclaurin theorem 
\begin{eqnarray}
\sum_{n=0}^N f(n)&=&\int_0^{N+1} f(n)\,\mathrm{d}n+\frac{1}{2}\left\{f(0)-f(N+1)\right\}\nonumber\\
&+&\sum_{n=1}^{\infty}\frac{B_{2n}}{(2n)!}\left\{f^{(2n-1)}(N+1)-f^{(2n-1)}(0)\right\},\nonumber\\
\end{eqnarray}
where the cutoff $N$ will be eventually sent to infinity. Let us apply the previous formula to $\Pi_{VV}$ and $\Pi_{TT}^-$\,:
\begin{eqnarray}\label{EMc_corr}
\Pi_{VV}(q^2)&=&\int_{n_{\Lambda}}^{N+1}\mathrm{d}n\, \frac{f_{Vn}^2}{-\,q^2+m_{Vn}^2}\,+\,\cdots\, ,\nonumber\\
\Pi_{TT}^-(q^2)&=&\int_{n_{\Lambda}}^{N+1}\,\mathrm{d}n\, \frac{(f_{Vn}^T)^2}{-\,q^2+m_{Vn}^2}\,+\,\cdots\, .
\end{eqnarray}
We will be interested in performing a large $q^2$ expansion. By using $n_{\Lambda}$ we have split the integrals keeping the contribution that will match the parton model logarithm of perturbative QCD [\,{\it{cf.}} Eqs.~(\ref{opevvtt})\,].\footnote{Recall that we are working at leading order in the strong coupling constant and, consequently, we are dismissing the fact that $\Pi_{TT}$ has a non-vanishing anomalous dimension.} The remaining piece, together with the omitted terms, only contribute as inverse powers of $q^2$. Their contribution determines the OPE condensates and is in general model dependent.

By looking at Eqs.~(\ref{EMc_corr}) one concludes that for highly excited resonances $f_{Vn}^{(T)\,2}=A_{V}^{(T)\,2} \frac{\mathrm{d}}{\mathrm{d}n}m_{Vn}^2$ for both vector and tensor decay constants. Actually this is the only possibility if we want to ensure the right high energy behavior. In other words, $f_{Vn}$ and $f_{Vn}^T$ have the same asymptotic $n$-scaling, regardless of the specific scaling the masses may take. Notice that this is only possible because $\Pi_{VV}$ and $\Pi_{TT}^-$ are both saturated by the exchange of $(1^{--})$ vector mesons. 

The scaling of the vector and tensor decay constants makes it possible to convert the integrals in
Eqs.~(\ref{EMc_corr}) over the radial excitation number $n$ into integrals over the 
mass. The integration is performed straightforwardly, yielding\,:
\begin{equation}
A_V^{(T)\,2}\!\!\int_{m_{n_{\Lambda}}^{2}}^{m_{N+1}^{2}}\!\!\mathrm{d}m^{2}\dfrac{m^{2}}{m^{2}-q^{2}}\,=\,A_V^{(T)\,2}\log\!\left(\dfrac{m_{N+1}^{2}-q^{2}}{m_{n_{\Lambda}}^{2}-q^{2}}\right)\!.
\end{equation}
It is important to stress that the limits $N\to\infty$ and $q^2\to\infty$ above do not commute. The
former must be taken in the first place, and together with the requirement 
$\lim_{n\to\infty}m_{n}=\infty$ the parton model logarithm is reproduced. 
Moreover, imposing that the quark-gluon picture is dual to the hadronic one, we obtain 
\begin{equation}\label{AA}
A_{V}^{2}\,=\,2\, A_{V}^{T\,2}\,=\,\dfrac{N_{c}}{12\,\pi^{2}}\,,
\end{equation}
and therefore
\begin{equation}\label{pred1}
\lim_{n\rightarrow \infty} \xi_n^2\,=\,\frac{A_{V}^{T\,2}}{A_V^2}=\frac{1}{2}\, ,
\end{equation}
where the constant is the ratio of the parton model coefficients for $\Pi_{VV}$ and $\Pi_{TT}^-$. 

Incidentally, note that in the previous result no use was made of the $b_1$ mesons entering $\Pi_{TT}^+$. In order to relate both parity sectors in $\Pi_{TT}$, additional assumptions on the spectrum would have to be made. For instance, if some relation between $m_{Vn}$ and $m_{Bn}$ were specified, a prediction for $f_{Vn}^T/f_{Bn}$ would then follow.

We now turn our attention to the crossed-correlator $\Pi_{VT}$. In this case, the Euler-Maclaurin theorem takes the form
\begin{equation}\label{EMVT}
\Pi_{VT}(q^{2})\,=\,\int_{n_{\Lambda}}^{N+1}\mathrm{d}n\,\xi_n\,\frac{f_{Vn}^{2}\,m_{Vn}}{-\,q^2+m_{Vn}^{2}}\,+\,\cdots\,.
\end{equation}
With the help of our previous combined analysis of $\Pi_{VV}$ and $\Pi_{TT}^-$, one concludes that the integrand diverges with an extra power of $m_{Vn}$. In a similar fashion as before, we can transform Eq.~(\ref{EMVT}) into an integral over the mass. Taking the cutoff to infinity, we would na\"\i vely obtain
\begin{equation}
\Pi_{VT}(q^2)\,=\,-\,\frac{N_c}{24\,\sqrt{2}}\sqrt{-q^2}\,+\,\cdots\, .
\end{equation}
However, in order to comply with the short distance behavior of Eq.~(\ref{shortvt}), it should {\it{converge}} as $q^{-2}$. The only possibility left is to allow for an alternate series, with $\xi_n$ showing a pattern of alternation in sign.\footnote{Note that $\xi_n$ is a real number because $f_V$ and $f_V^T$ are defined to be real, so this is indeed the only possible scenario. To the best of our knowledge, the first instance of alternating contributions in the hadronic spectrum was found in Ref.~\cite{Bramon:72} in the context of $e^+e^-\rightarrow$ hadrons.} Notice that this does not pose any problem\,: only the magnitude of $\xi_n$ was determined in Eq.~(\ref{pred1}). Note also that, unlike $\Pi_{VV}$ and $\Pi_{TT}^-$, $\Pi_{VT}$ is not positive definite and in principle \textbf{it} can contain both positive and negative contributions. 

The most general situation that complies with QCD is the presence of some cancellations for high-resonance contributions, no matter how they are arranged. The simplest (and most natural) scenario consists of a regular pattern of sign-alternating contributions. For this particular scenario, consistency with perturbative QCD leads to the prediction
\begin{equation}\label{pred}
\xi_n\,=\,(-1)^n\,|\xi_n|\,,\qquad |\xi_n|\,=\,\frac{1}{\sqrt{2}}\,\simeq\, 0.71\, ,
\end{equation}
for highly excited $\rho$-like vector meson resonances. Interestingly, $\xi_{\rho}$ has been recently computed in the lattice~\cite{Bec,Braun:2003jg,Donnellan:2007xr}. Quite remarkably, the value reported is $\xi_{\rho}(2\,{\mathrm{GeV}})=0.72(2)$ at $\mu=2$ GeV, in excellent agreement with our asymptotic large-$N_c$ result.\footnote{In a recent paper~\cite{Mateu:2007tr} this ratio was also determined for $\mu=1\,\mathrm{GeV}$, the value reported being $\xi_{\rho}=0.75(14)$. Sum rules also obtain similar results~\cite{Ball:2002ps,Bakulev:1999gf}. Incidentally, in the ENJL model~\cite{Chizhov:2003qy} one also finds $\xi_{\rho}=1/\sqrt{2}$.}

This result is extremely interesting, suggesting that $\xi_n$ may be a constant independent of the resonance excitation number. Therefore, one would like to assess what is the range of validity of the pattern shown in Eq.~(\ref{pred}). Incidentally, one would also like to identify the specific realization of opposite-sign contributions. It would certainly look odd if the alternation started at some energy scale $\mu\sim m_{\rho_{\bar{n}}}$, but it cannot be ruled out. However, if this were the case, some triggering dynamical mechanism at this scale should be invoked. The natural thing to expect is that a regular pattern of sign-flipping contributions be a feature of the whole meson tower.

So far we have been dealing with large-$N_c$ QCD. A more ambitious and interesting issue is to check whether the result of Eq.~(\ref{pred}) and the conjectured opposite-sign pattern we advocate as its most natural realization  has anything to do with QCD. In the following section we will see that QCD finite energy sum rules nicely comply with this picture. 

\section{Comparison with QCD spectral sum rules}
\label{sec:sumrules}

\noindent In order to test the ideas of the previous section, we will consider a set of sum rules. We will start with the $\Pi_{VV}$ and $\Pi_{TT}^-$ correlators and afterwards consider $\Pi_{VT}$.

We choose as hadronic {\it{ans\"atze}} the following functions,
\begin{eqnarray}
\frac{1}{\pi}\,{\mathrm{Im}}\,\Pi_{VV}(t)&=&f_{\rho}^2\,\delta(t-m_{\rho}^2)\,+\,f_{\rho\prime}^2\,\delta(t-m_{\rho\prime}^2)\nonumber\\
&\quad+&\frac{4}{3}\frac{N_c}{(4\,\pi)^2}\,\kappa_V\,\theta(t-s_0)\, ,\nonumber\\
\frac{1}{\pi}\,{\mathrm{Im}}\,\Pi_{TT}^-(t)&=&\xi_{\rho}^2\,f_{\rho}^2\,\delta(t-m_{\rho}^2)\,+\,
\xi_{\rho\prime}^2\,f_{\rho\prime}^2\,\delta(t-m_{\rho\prime}^2)\nonumber\\
&\quad+\,&\frac{2}{3}\frac{N_c}{(4\,\pi)^2}\,\kappa_T\,\theta(t-{\bar{s}}_0)\, ,
\end{eqnarray}
consisting of two isolated single poles, corresponding to the $\rho(770)$ and $\rho(1450)$ plus a continuum, whose onset is determined by the parameters $s_0$ and ${\bar{s}}_0$, which in general are different. The factors in front of the theta terms have been chosen so as to match the parton model logarithms of Eqs.~(\ref{opevvtt}). The parameters $\kappa_{T}$, $\kappa_{V}$ are given by
\begin{eqnarray}
\kappa_T(\mu)&=&1\,+\,\frac{\alpha_s(\mu)}{3\,\pi}\left(\frac{7}{3}\,+\,2\,\log\frac{t}{\mu^2}\right)\,,\\
\kappa_V&=&1\,+\,\frac{\alpha_s(\mu)}{\pi}\,.\nonumber
\end{eqnarray}
They represent the first-order $\alpha_s$ correction to the perturbative contribution, the former also accounting for the fact that the tensor current has a non-vanishing anomalous dimension.

Using the dispersion relations of Eqs.~(\ref{disprel}), expanding the result in inverse powers of momenta and matching onto the short distance results of Eqs.~(\ref{opevvtt}), one finds
\begin{eqnarray}\label{1}
f_{\rho}^2+f_{\rho\prime}^2-\frac{4}{3}\frac{N_c}{(4\,\pi)^2}\,\kappa_V\,s_0&=&0\, ,\nonumber\\
\xi_{\rho}^2\,f_{\rho}^2+\xi_{\rho\prime}^2\,f_{\rho\prime}^2
+\Lambda_3-\frac{2}{3}\frac{N_c}{(4\,\pi)^2}\,\kappa_2\,{\bar{s}}_0&=&0\, ,\nonumber\\
f_{\rho}^2\,m_{\rho}^2+f_{\rho\prime}^2\,m_{\rho\prime}^2-\frac{2}{3}\frac{N_c}{(4\,\pi)^2}\,\kappa_V\,s_0^2&=&\nn
-\,\frac{1}{12\,\pi}\,\langle \alpha_s G^{\mu\nu}G_{\mu\nu}\rangle&&\phantom{0},\nonumber\\
\xi_{\rho}^2\,f_{\rho}^2\,m_{\rho}^2+\xi_{\rho\prime}^2\,f_{\rho\prime}^2\,m_{\rho\prime}^2-\frac{1}{3}\frac{N_c}{(4\,\pi)^2}\,\kappa_4\,{\bar{s}}_0^2&=&\nn
\frac{1}{24\,\pi}\,\langle \alpha_s G^{\mu\nu}G_{\mu\nu}\rangle&&\phantom{0},
\end{eqnarray}
where $\kappa_2$ and $\kappa_4$ are given by
\begin{eqnarray}\label{corrtensor}
\kappa_2({\bar{s}}_0)&=&1+\frac{1}{9}\frac{\alpha_s(\sqrt{{\bar{s}}_0})}{\pi}\, ,\nonumber\\
\kappa_4({\bar{s}}_0)&=&1+\frac{4}{9}\frac{\alpha_s(\sqrt{{\bar{s}}_0})}{\pi}\, .
\end{eqnarray}
Notice that above the renormalization point was chosen to be $\mu^2={\bar{s}}_0$.

As already noticed in Ref.~\cite{7-11}, $\alpha_s$ corrections in the vector channel induce at most a $8\%$ change in the decay constants and will be dismissed. For the tensor channel, the equations above show that the $\alpha_s$ correction in the sum rules is extremely small. For instance, at $\bar{s}_0=1.5$ GeV$^2$, they represent less than 2$\%$ for $\kappa_2$ and about 6$\%$ for $\kappa_4$. Therefore, the perturbative corrections in $\alpha_s$ can be safely neglected. 

For the numerical analysis, we will take as inputs the masses, $m_{\rho}=770$ MeV and $m_{\rho\prime}=1440$ MeV, and the gluon condensate. Due to the existing uncertainty, we will choose it to lay in the range $\langle \alpha_s G^{\mu\nu}G_{\mu\nu}\rangle=(0.001-0.021)\,\pi$ GeV$^4$, which includes both the values extracted from charmonium sum rules and $\tau$ decays~\cite{sumrules}. Additionally, we will use the relation $f_\rho=\sqrt{2}\,f_\pi$, which comes from assuming unsubtracted dispersion relations for both the pion electromagnetic form factor and the axial form factor in radiative pion decay~\cite{Ecker} and has been shown to be satisfied in sum rule analysis of vector and axial channels. With $f_{\pi}=131$ MeV, one obtains $f_{\rho}=185$ MeV. Finally, we will further impose $\xi_{\rho}^2=0.5$, in accord with the lattice determination. Notice that in the sum rules we have included the $\Lambda_3$ term. However, lacking any estimate of the parameter, for our numerical analysis we will set $\Lambda_3=0$, as commonly assumed in the literature.

Solving Eqs.~(\ref{1}) for $f_{\rho\prime}$, $s_0$, ${\bar{s}}_0$ and $\xi_{\rho\prime}$, one finds
\begin{eqnarray}
\sqrt{s_0}=(1.64\pm0.02)\,{\mathrm{GeV}}\, ,
&&\sqrt{{\bar{s}}_0}=(1.59\pm0.02)\,{\mathrm{GeV}}\,,\nonumber\\
f_{\rho\prime}=(182\pm 5)\,{\mathrm{MeV}}\,,&&\xi_{\rho\prime}=(0.95\pm 0.05)\, \xi_{\rho}\, ,
\end{eqnarray} 
where the errors quoted are due to the variation of the gluon condensate. Note that both $s_0$ and ${\bar{s}}_0$ yield reasonable values, {\it{i.e.}}, they satisfy $m_{\rho(1440)}<\sqrt{s_0}\sim \sqrt{{\bar{s}}_0}< m_{\rho(1750)}$.\vspace*{1.5mm}

The following comments are in order\,:
\begin{itemize}
   \item Lower values of the gluon condensate, typical in analysis of $\tau$ decays, favor $\xi_{\rho\prime}\sim \xi_{\rho}$. In particular, notice that a vanishing gluon condensate, not excluded by $\tau$ decay analyses, implies $\xi_{\rho\prime}=\xi_{\rho}$ (together with $s_0={\bar{s}}_0$).
   \item Eqs.~(\ref{1}) provide a solution only for the narrow range $178$ MeV $\leq f_{\rho}\leq 188$ MeV. Interestingly, the range complies with the relation $f_{\rho}^2\simeq2f_{\pi}^2$.
\end{itemize}

In order to test our conjectured pattern of signs we have to consider $\Pi_{VT}$.  Our spectral {\it{ansatz}} will be the following\,:
\begin{equation}\label{ansatzVT}
\frac{1}{\pi}\,{\mathrm{Im}}\,\Pi_{VT}(t)=\xi_{\rho}f_{\rho}^2m_{\rho}\delta(t-m_{\rho}^2)+\xi_{\rho\prime}f_{\rho\prime}^2m_{\rho\prime}\delta(t-m_{\rho\prime}^2).
\end{equation}
Inserting the last expression and the OPE of Eq.~(\ref{shortvt}) into the dispersion relation and equating powers of $q^2$ on both sides we get
\begin{eqnarray}
-\,2\,\langle {\bar{\psi}} \psi \rangle&=&\xi_{\rho}\,f_{\rho}^2\,m_{\rho}\,+\,\xi_{\rho\prime}\,f_{\rho\prime}^2\,m_{\rho\prime}\, ,\nonumber\\
-\,\frac{2\,g_s}{3}\,\langle {\bar{\psi}}\,\sigma_{\mu\nu}\,G^{\mu\nu}\psi\rangle&=&\xi_{\rho}\,f_{\rho}^2\,m_{\rho}^3\,+\,\xi_{\rho\prime}\,f_{\rho\prime}^2\,m_{\rho\prime}^3\, .
\end{eqnarray}
Upon solving these equations for $\xi_{\rho}$ and $\xi_{\rho^{\prime}}$ we find
\begin{eqnarray}\label{solSR}
\xi_{\rho\prime}&=&\frac{2\,\langle{\bar{\psi}}\psi\,\rangle\, m_{\rho}^2}{f_{\rho\prime}^2\, m_{\rho\prime}\,(m_{\rho\prime}^2-m_{\rho}^2)}\left[1-\frac{\lambda}{m_{\rho}^2}\right]\, ,\nonumber\\
\xi_{\rho}&=&-\,\frac{2\,\langle{\bar{\psi}}\psi\rangle \,m_{\rho\prime}^2}{f_{\rho}^2\, m_{\rho}\,(m_{\rho\prime}^2-m_{\rho}^2)}\left[1-\frac{\lambda}{m_{\rho\prime}^2}\right]\, ,
\end{eqnarray}
where 
\begin{equation}
\lambda\,=\,\frac{g_s}{3}\,\frac{\langle {\bar{\psi}}\,\sigma_{\mu\nu}\,G^{\mu\nu}\psi\rangle}{\langle{\bar{\psi}}\psi\rangle}\,,
\end{equation}
is the ratio between the mixed and the quark condensate. In view of Eqs.~(\ref{solSR}) there are three possible scenarios, depending on the magnitude of $\lambda$ (recall that the quark condensate is negative)\,:
\begin{itemize} 
     \item $\lambda<m_{\rho}^2$, leading to alternation in sign, with positive $\xi_{\rho}$;
     \item $m_{\rho}^2<\lambda<m_{\rho\prime}^2$, where both $\xi_{\rho}$ and $\xi_{\rho\prime}$ are positive; 
     \item $\lambda>m_{\rho\prime}^2$, leading to alternation in sign but with a negative $\xi_{\rho}$. 
\end{itemize}
The last possibility is in clear contradiction with the lattice result and can be readily excluded. Independent sum rule analyses indeed concluded that~\cite{Belyaev:82}
\begin{equation}
\lambda \,\sim\, 0.22\, {\mathrm{GeV^2}}\,<\, m_{\rho}^2\, ,
\end{equation}
so that the mixed condensate is small enough and leads to alternation in sign.\footnote{Incidentally, notice that arbitrarily large negative values of $\lambda$ would have also led to this scenario.} Note that the small value of the mixed condensate in the second equation forces the alternation in sign, whereas the quark condensate fixes the contribution of the $\rho(770)$ to be positive.

More sophisticated sum rules have confirmed the pattern of alternating contributions in $\Pi_{VT}$~\cite{Bal}. However, a word of caution should be issued on the quantitative values of the parameters extracted from such sum rules. We already pointed out in the previous section that the presence of a mass factor multiplying each resonance contribution in Eq.~(\ref{ansatzVT}) spoils the convergence of the series. As a result, the sum rules are not stable under addition of new resonance states in the spectral function. However, Eqs.~(\ref{solSR}) distinctly show that there has to be some negative contribution in the spectrum of $\Pi_{VT}$ to outweight the $\rho(770)$ contribution.  

\section{Discussion}
\label{sec:disc}

\noindent A remarkable property of QCD in the large-$N_c$ limit is that the qualitative characteristics of hadrons emerge naturally from imposing quark-hadron duality consistency conditions on the correlators of the theory. This very general analysis does not rely on the particular flavor or Dirac structure of the correlators. Therefore, any relation between a certain subset of correlators may turn out to yield additional useful constraints on the spectrum of large-$N_c$ QCD. 

In this paper we have shown that, for the set of correlators $\Pi_{VV}$, $\Pi_{TT}^-$ and $\Pi_{VT}$, even quantitative predictions can be extracted. From a combined analysis  we concluded that, for highly excited states, $f_{Vn}^T/f_{Vn}\sim (-1)^n|f_{Vn}^T/f_{Vn}|$, where $|f_{Vn}^T/f_{Vn}|=1/\sqrt{2}$. The ratio of decay constants is fixed by the Dirac structure of the currents and equals the ratio of the leading perturbative terms of $\Pi_{VV}$ and $\Pi_{TT}$, while the alternation in sign is required to ensure the convergence of $\Pi_{VT}$.

We find this result particularly beautiful. It is a really striking prediction which relies only on the simultaneous high-energy consistency of the correlators. In this sense, the previous result can be rendered as a high energy theorem of large-$N_c$ QCD. Our analysis was restricted to light-flavor vector mesons, but similar predictions should be obtained for mesons with heavy flavors.

A natural issue to address at this point is whether this pattern, valid for highly excited mesons in the large-$N_c$ limit, resembles QCD. The lattice recently computed the ratio of the $\rho(770)$ decay constants, with the result $f_{\rho}^T/f_{\rho}=0.72(2)$. The agreement is certainly impressive, and it seems suggestive to entertain the scenario of $n$-independent decay constant ratios for the $\rho$-meson radial excitations. We tested this possibility with QCD sum rules and the pattern is reproduced to a remarkable degree, specially for low values of the gluon condensate.

In this paper we have concentrated on the high energy behavior of the ``bootstrap" correlators. It would be very interesting to investigate the consequences that our results shed at low energies. In a recent paper we already worked out the low energy effective field theory of chirally-symmetric operators coupled to tensor sources~\cite{CataMateu07}. An interesting issue one can now address is the impact of Eqs.~(\ref{pred}) on the predictions for the low energy chiral couplings.   
 
\begin{acknowledgments}
\noindent O.~C. wants to thank S.~R.~Sharpe for valuable discussions at the different stages of this work and to A.~Bramon for pointing out the existence of a similar opposite-sign pattern in $e^+e^-\rightarrow$ hadrons. V.~M. wants to thank the hospitality of U.~C.~San Diego, where part of this work was performed. V.~M. also thanks Professor J.~Novotn\'y for valuable discussions on $\Lambda_3$. The work of O.~C. is supported by the Fulbright Program and the Spanish Ministry of Education and Science under grant no. FU2005-0791. The work of V.~M. is supported by a FPU contract (MEC), by EU MRTN-CT-2006-035482 (FLAVIAnet), MEC (Spain) under grant FPA2004-00996 and by Generalitat Valenciana under grants GRUPOS03/013 and GV05/164.
\end{acknowledgments}



\begin{thebibliography}{99}

\bibitem{'tHooft:73}
  G.~'t Hooft,
  Nucl.\ Phys.\  B {\bf 72} (1974) 461.

\bibitem{Ecker}
  G.~Ecker, J.~Gasser, H.~Leutwyler, A.~Pich and E.~de Rafael,
  Phys.\ Lett.\  B {\bf 223} (1989) 425.
  
\bibitem{Witten:79}
  E.~Witten,
  Nucl.\ Phys.\  B {\bf 160} (1979) 57.
  
\bibitem{rev}
  E.~De Rafael,
  AIP Conf.\ Proc.\  {\bf 602} (2001) 14;
  M.~Knecht, S.~Peris and E.~de Rafael,
  Nucl.\ Phys.\ Proc.\ Suppl.\  {\bf 86} (2000) 279.

\bibitem{works}
  See, for instance, M.~Knecht, S.~Peris and E.~de Rafael,
  Phys.\ Lett.\  B {\bf 508} (2001) 117;
  O.~Cata and S.~Peris,
  JHEP {\bf 0407} (2004) 079;
  S.~Peris and E.~de Rafael,
  Phys.\ Lett.\  B {\bf 490} (2000) 213;
  O.~Cata and S.~Peris,
  JHEP {\bf 0303} (2003) 060;
  T.~Hambye, S.~Peris and E.~de Rafael,
  JHEP {\bf 0305} (2003) 027;
  M.~Knecht, S.~Peris, M.~Perrottet and E.~De Rafael,
  JHEP {\bf 0211} (2002) 003;
  M.~Knecht, A.~Nyffeler, M.~Perrottet and E.~De Rafael,
  Phys.\ Rev.\ Lett.\  {\bf 88} (2002) 071802;
  M.~Knecht, S.~Peris and E.~de Rafael,
  Phys.\ Lett.\  B {\bf 443} (1998) 255.

\bibitem{Craigie:81}
  N.~S.~Craigie and J.~Stern,
  Phys.\ Rev.\  D {\bf 26}, 2430 (1982).
  

\bibitem{Bec}
  D.~Becirevic, V.~Lubicz, F.~Mescia and C.~Tarantino,
  JHEP {\bf 0305} (2003) 007.
 
\bibitem{Braun:2003jg}
  V.~M.~Braun, T.~Burch, C.~Gattringer, M.~Gockeler, G.~Lacagnina, S.~Schaefer and A.~Schafer,
  Phys.\ Rev.\  D {\bf 68} (2003) 054501.

\bibitem{Donnellan:2007xr}
  M.~A.~Donnellan {\it et al.},
  [arXiv:0710.0869 [hep-lat]].
  
\bibitem{CataMateu07}
  O.~Cata and V.~Mateu,
  JHEP {\bf 0709} (2007) 078.



\bibitem{Gov}
  J.~Govaerts, L.~J.~Reinders, F.~de Viron and J.~Weyers,
  Nucl.\ Phys.\ B {\bf 283} (1987) 706.

\bibitem{Bal}
  I.~I.~Balitsky, A.~V.~Kolesnichenko and A.~V.~Yung,
  Sov.\ J.\ Nucl.\ Phys.\  {\bf 41} (1985) 178
  [Yad.\ Fiz.\  {\bf 41} (1985) 282].
  
\bibitem{Bramon:72}
  A.~Bramon, E.~Etim and M.~Greco,
  Phys.\ Lett.\  B {\bf 41}, 609 (1972).

\bibitem{Mateu:2007tr}
  V.~Mateu and J.~Portoles,
  Eur.\ Phys.\ J.\  C {\bf 52} (2007) 325.

\bibitem{Ball:2002ps}
  P.~Ball and V.~M.~Braun,
  Phys.\ Rev.\  D {\bf 54} (1996) 2182;
  P.~Ball, V.~M.~Braun and N.~Kivel,
  Nucl.\ Phys.\  B {\bf 649} (2003) 263.

\bibitem{Bakulev:1999gf}
  A.~P.~Bakulev and S.~V.~Mikhailov,
  Eur.\ Phys.\ J.\  C {\bf 17} (2000) 129.

\bibitem{Chizhov:2003qy}
  M.~V.~Chizhov,
  JETP Lett.\  {\bf 80} (2004) 73
  [Pisma Zh.\ Eksp.\ Teor.\ Fiz.\  {\bf 80} (2004) 81].
  
\bibitem{7-11}
  M.~F.~L.~Golterman and S.~Peris,
  Phys.\ Rev.\  D {\bf 61} (2000) 034018.
  
\bibitem{sumrules}
  See, for instance, M.~Davier, A.~Hocker and Z.~Zhang,
  Nucl.\ Phys.\ Proc.\ Suppl.\  {\bf 169} (2007) 22;
B.~L.~Ioffe and K.~N.~Zyablyuk,
  Eur.\ Phys.\ J.\  C {\bf 27} (2003) 229;
F.~J.~Yndurain,
  Phys.\ Rept.\  {\bf 320} (1999) 287.

  

\bibitem{Belyaev:82}
  V.~M.~Belyaev and B.~L.~Ioffe,
  Sov.\ Phys.\ JETP {\bf 56} (1982) 493
  [Zh.\ Eksp.\ Teor.\ Fiz.\  {\bf 83} (1982) 876].






\end{thebibliography}
\end{document}